# In-Process Monitoring of Gear Power Honing Using Vibration Signal Analysis and Machine Learning

Massimo Capurso[a], Luciano Afferrante[a],*

[a] Department of Mechanics, Mathematics and Management, Polytechnic University of Bari, Via E. Orabona 4, Bari, 70125, Italy

* Corresponding Author: E-mail: luciano.afferrante@poliba.it

**ABSTRACT**

In modern gear manufacturing, stringent Noise, Vibration, and Harshness (NVH) requirements demand high-precision finishing operations such as power honing. Conventional quality control strategies rely on post-process inspections and Statistical Process Control (SPC), which fail to capture transient machining anomalies and cannot ensure real-time defect detection. This study proposes a novel, data-driven framework for in-process monitoring of gear power honing using vibration signal analysis and machine learning. Our proposed methodology involves continuous data acquisition via accelerometers, followed by sophisticated time-frequency signal analysis to extract discriminative features. We systematically investigate and compare the efficacy of three subspace learning methods for feature extraction: (1) Principal Component Analysis (PCA) for dimensionality reduction; (2) a two-stage framework combining PCA with Linear Discriminant Analysis (LDA) for enhanced class separation; and (3) Uncorrelated Multilinear Discriminant Analysis with Regularization (R-UMLDA), a multilinear extension of LDA specifically adapted for tensor data, which enforces feature decorrelation and includes regularization for small sample sizes. These extracted features are then fed into a Support Vector Machine (SVM) classifier to predict four distinct gear quality categories, established through rigorous geometrical inspections and test bench results of assembled gearboxes. The models are trained and validated on an experimental dataset collected in an industrial context during gear power-honing operations, with gears classified into four different quality categories, defined according gears geometrical inspections and test bench results on the assembled gearboxes. The proposed framework achieves high classification accuracy (up to 100%) in an industrial setting, validating its robustness. Beyond classification, the approach offers interpretable spectral features that correlate with process dynamics, enabling practical integration into real-time monitoring and predictive maintenance systems.

**Keywords:** In-Process Monitoring, Vibration Signal Analysis, Subspace Learning, Smart Manufacturing, Support Vector Machine (SVM) Classification, Predictive Maintenance.

# 1 Introduction

## 1.1 Context and Motivation

The automotive industry faces increasingly stringent Noise, Vibration, and Harshness (NVH) requirements, demanding superior vibro-acoustic performance from newly manufactured transmissions [1]. Currently, the fulfillment of these requirements is predominantly assessed through time-consuming and costly End-of-Line (EOL) tests conducted on assembled gearboxes. A critical drawback of EOL systems is their limited integration with process improvement cycles, as they provide insufficient real-time feedback to identify and mitigate the root causes of non-conformities, leading to substantial economic losses. Manufacturing defects and suboptimal gear geometry are primary contributors to noise and vibration in transmissions [2]. To address these challenges, hard-finishing processes, such as continuous generative grinding and honing, are employed at the end of the gear production chain. These processes, characterized by undefined cutting edges, are crucial for mitigating thermal distortions, achieving micro-geometric corrections, and enhancing surface quality on tooth flanks, thereby meeting stringent quality and vibro-acoustic standards [3]. Gear Power Honing, in particular, has emerged over the last two decades as a cost-effective and competitive alternative to grinding, yielding gears with low-noise running characteristics due to the absence of parallel exciting structures on the tooth flank [4]. However, gear honing is susceptible to dynamic instability phenomena, such as chatter or self-excited vibrations, primarily due to variable contact conditions and variations in the quality of raw input parts [5]. These instabilities can induce periodic structures or waviness on tooth surfaces, significantly contributing to transmission error and generating undesirable vibration and noise in gear systems [6]. These defects manifest as distinct peaks, known as "ghost components," in the vibration spectrum at frequencies related to the wavelength and direction of the tooth flank waviness [7], [8], [9].

## 1.2 State of Art in Process Monitoring

Significant advancements have been made in machine tool design and gear manufacturing to meet evolving quality demands. While structural optimizations and advanced damping systems have been introduced in grinding machines to reduce dynamic instabilities, the detection of surface waviness and prediction of vibro-acoustic behavior often still rely on expensive and time-intensive investigation systems. Statistical Process Control (SPC) methods are commonly applied in production, typically after machine downtime or tool changes, to ensure final quality and adjust process parameters. However, SPC methods are inherently limited in their ability to capture transient phenomena that can compromise final product quality during continuous operation. Evaluating gear geometry comprehensively, as per VDI 2612 standards including ripple evaluation [10] requires assessing the topography of all teeth, a practice hindered by high production rates in industrial settings. Although inline inspection systems utilizing roll testing and laser scanning have been developed [11], they necessitate substantial investment in advanced measurement equipment. In-process monitoring systems, which utilize continuous data acquisition from sensors mounted on machine tools, offer a robust, cost-effective, and flexible solution that complements traditional SPC methods and can be readily implemented on existing machinery [12]. Despite their promise, accurately predicting the quality of finished components from real-time process data remains a significant research challenge [13]. Current in-process monitoring solutions for finishing machines largely rely on amplitude control of process quantities such as power absorption and acceleration [14]. Specifically, acceleration monitoring is employed to detect forced vibrations from internal or external sources and to identify self-excited vibrations [15]. Given that gear noise is often attributed to dynamic periodic phenomena during machining [8], spectrum masks applied to acceleration frequency spectra are a common control criterion [16]. However, the effectiveness of this approach in machining processes is often compromised by fluctuating variables like tool wear, workpiece input quality, temperature, and process parameters (e.g., tool dressing cycles, feed rate), all of which can affect signal amplitude and frequency. This variability makes the setting of fixed thresholds challenging, often requiring arduous trial-and-error methods and continuous manual adjustment, thereby rendering monitoring operations time-consuming and heavily reliant on human expertise. In the field of process monitoring for continuous generative grinding, several signal processing approaches have been proposed in the literature to identify process anomalies and detect substandard components during machining. Böttger J. et al. [17] proposed a monitoring method using current and acoustic emission sensors to assess abrasive tool condition and workpiece quality, linking process deviations to vibro-acoustic parameters through simulation. The analysis of these signals makes possible to evaluate the condition of the abrasive tool and the quality of the workpiece, optimizing process parameters to reduce cycle time and extend tool life. In addition, through the grinding process simulation model, the relationship between the process deviations and the relevant vibro-acoustic parameters on tooth surface can be identified [18]. Data-driven models based on Convolutional Neural Networks (CNNs) have also shown promise for in-process monitoring in grinding, with Liu et al. employing CNNs to classify machine faults based on time-frequency representations (Short-Time Fourier Transform - STFT or Wavelet Packet Decomposition - WPD) of vibration signals [19].

In the context of data-driven process and condition-based monitoring, subspace learning techniques have gained significant traction. Principal Component Analysis (PCA) is widely used for dimensionality reduction in high-dimensional and correlated datasets, aiming to obtain a reduced yet meaningful variable space [20], [21]. Its utility in image processing for dimensionality reduction and feature extraction, leading to "Eigenfaces," was first demonstrated by Sirovich and Kirby in Ref. [22] , and later applied to face recognition by

Turk and Pentland [23]. To enhance classification performance, particularly in challenging real-world conditions, hybrid approaches like the "Fisher-faces" method proposed by Belhumeur et al. [24] proposed a combined approach based on PCA and LDA, known as Fisher-face [25] and for bearing fault diagnosis using vibration spectrograms, where principal components (Eigen-spectrograms) effectively capture underlying patterns for classification [26].

More advanced multilinear subspace learning methods, such as Uncorrelated Multilinear Discriminant Analysis (UMLDA), directly operate on high-order tensor data, preserving their natural structure by performing mode-wise projections. Lu et al. [27] and [28] extended classical LDA to tensor data by introducing discriminant analysis along multiple modes and incorporating a regularization term to mitigate the small sample size problem and avoid singular scatter matrices. Ye et al. [29], further improved UMLDA for multi-channel profile data analysis, demonstrating its effectiveness in detecting process deviations and identifying fault types for high-precision industrial applications.

## 1.3  Current work and Contribution

Despite the growing interest in data-driven process monitoring techniques, their specific application to in-process quality control in gear honing remains largely underexplored in literature. This work proposes a predictive, AI-integrated strategy for gear quality assessment, based on time-frequency analysis of multi-channel vibration signals acquired directly during the gear power honing process. Our monitoring approach utilizes signals from two accelerometers, which are processed using the Short-Time Fourier Transform (STFT) to generate time-frequency spectrograms. A key innovative aspect of our pre-processing involves combining these two spectrograms into a single bi-dimensional input map for each observation via element-wise multiplication, aiming to emphasize shared features and reduce sensor-specific noise. To effectively reduce dimensionality and extract highly informative features, these spectrograms are projected into a lower-dimensional subspace defined by a set of "eigen-maps". These eigen-maps are designed to capture the dominant variance and identify distinct process conditions. We comprehensively explore and compare the performance of three state-of-the-art subspace learning methods: (1) Principal Component Analysis (PCA), applied following the methodology in Ref. [26]; (2) PCA combined with Linear Discriminant Analysis (LDA), inspired by the "Fisher-faces" approach [24]; (3) Uncorrelated Multilinear Discriminant Analysis with Regularization (R-UMLDA), as detailed in [27], which is applied to both the combined bi-dimensional maps and raw spectrogram pairs (third-order tensors) to assess the impact of data structure preservation. The resulting projections in the eigen-map space are used as input for a kernelized multiclass SVM classifier to predict four gear quality classes.

Our experimental results demonstrate that vibration spectrograms can effectively and accurately predict final gear quality. This confirms and extends prior evidence on the reliability of vibration-based monitoring, particularly from grinding processes to the under-investigated domain of gear power honing. A significant contribution is the observation that the generated eigen-maps not only provide a low-dimensional representation but also reveal specific non-stationary phenomena in the time-frequency domain. These phenomena are directly linked to surface waviness on gear flanks, a known critical source of noise and vibration in transmission operation. These findings align with previous experimental and simulation-based studies on grinding dynamics and, collectively, underscore the suitability of the proposed methods for robust, real-time in-process monitoring in industrial environments. The successful application of R-UMLDA to raw tensor data (spectrogram pairs) without requiring prior flattening or merging further highlights the method's efficiency and potential for directly leveraging the inherent multi-dimensional structure of sensor data, which represents a notable advancement for industrial monitoring systems.

# 2 Case study

## 2.1 Process and acquisition set-up

The main components of a power honing machine are the honing head, workpiece spindle and tailstock as described in Figure 1. The workpiece and tool synchronously rotate at a constant speed and an oscillating shift in the axial direction is superimposed to the radial infeed to improve the final surface quality. In addition, the workpiece and honing tool axis are skewed under a cross-axis angle, which generate the relative movement necessary for the abrasive action. The resulting surface pattern led to low noise transmission [30]. The honing process is developed in three main phases: roughing phase characterized by a high radial infeed during which much of the allowance is removed, the finishing phase with reduced infeed to achieve the desired dimensional accuracy and spark-out phase with no radial infeed to improve surface quality. The vibrational signal is acquired by MEMS accelerometers with a sampling rate of 25 kHz, mounted on the spindle and on the tailstock (Figure 1). The time signal is processed by a control unit that performs a STFT with a 512-point Hamming window. The observation window was reduced by approximately 40% of the cycle time, removing the initial roughing step as it primarily reflects the component's initial quality. The resulting spectrograms contain $w = 200$ timeframe and $b = 512$ frequency bins. An on-board PC collects machine parameters, timestamps and the resulting spectrograms in a dedicated cloud system.

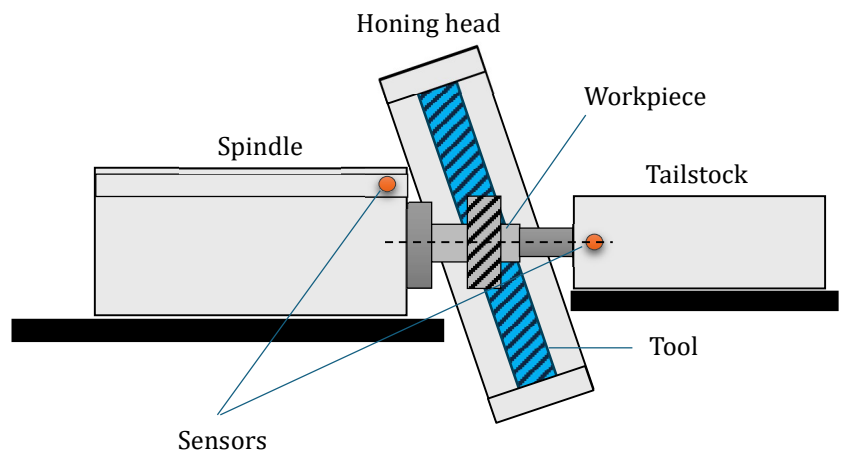

*Figure 1 - Conceptual drawing of the power honing machine and sensors layout.*

## 2.2 Experimental campaign

An experimental test campaign was carried out at an automotive transmission production facility, to collect real data during the gear power honing process. The machined component is a case-hardened pinion for dual-clutch transmission. The test campaign covered an observation period of about one year of production on a gear Power Honing machine, during which, according to the SPC protocol, 1% of the machined components were sampled. The sampled components were first measured to evaluate microgeometry and waviness deviations. A representative fraction of them, with different quality levels, was identified and collected and the associated vibration signal acquired during machining was recorded. At the end of the acquisition phase, pinions were assembled in the gearbox and tested to the EoL test bench. To ensure representative results during testing, the driven gears, in the meshing pair, were measured according to VDI 2612. During testing, compliance with quality control parameters was evaluated in terms of micro-geometry and NVH behavior. Based on these assessments, a subset of 429 components was selected and categorized into four distinct classes: (1) components with compliant quality in micro-geometry and vibro-acoustic terms (identified as OK), (2) components with non-compliant NVH performance in the EoL test, caused by surface waviness on gear flanks (identified as NOK 1), (3) components exhibiting microgeometry deviations on gear flanks while maintaining compliant NVH behavior (identified as NOK 2), and (4) a smaller group of components with non-compliant NVH behavior attributed to periodic pitch patterns on gear teeth (identified as NOK 3).

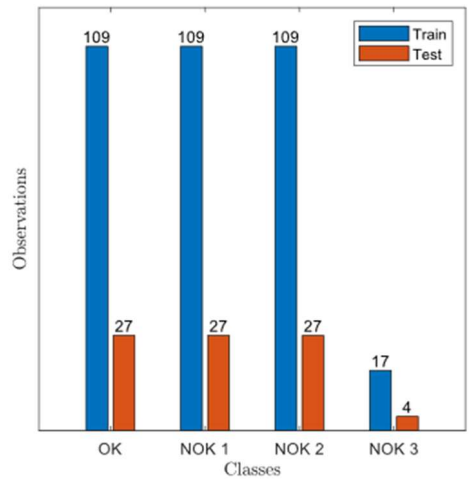

*Figure 2 - Dataset partitioning for each class.*

The dataset was then randomly partitioned into a training set (80%) and a test set (20%) to evaluate model performance. Figure 2 illustrates the class-wise distribution of observations within the training and test sets. Let's observe that the NOK3 class is notably represented by a lower number of samples, compared to the others. This condition allows for the evaluation of model performance under realistic industrial scenarios, where class imbalance is a common issue.

## 2.3 Methodology

Experimental evidence indicates that self-excited vibrations, occurring during both roughing and finishing phases, can be triggered at frequencies that are multiples of the workpiece's rotational frequency. The ultimate impact on gear quality is largely determined by the final stage of machining, where these vibrations may either intensify or rapidly decay. Critically, the influence of these phenomena on component quality is more effectively assessed by analyzing amplitude trends across individual frequency components rather than relying on global maximum amplitude values, which are frequently affected by external variables. This inherent complexity limits the effectiveness of traditional monitoring approaches based solely on amplitude thresholds in detecting critical process deviations. Moreover, depending on their frequency, these vibrations can adversely affect the pitch pattern between teeth or the surface finish of the tooth flanks. Therefore, to reliably characterize gear manufacturing defects from vibration data, it is imperative to retain the complete information embedded within time-frequency spectrograms, as they provide a comprehensive representation of the underlying process dynamics.

### 2.3.1 Spectrograms pre-processing

Signal amplitude is significantly influenced by various machine-related variables, predominantly tool wear and dressing conditions. To ensure consistency across fluctuating machine conditions and to balance the contribution from each sensor, all spectrograms are amplitude-normalized to a grayscale range of [0, 255]. While the acquired vibration signal reflects the dynamic behavior of the entire mechanical system, only a specific portion of this information is directly pertinent to the phenomena affecting the final gear quality. Furthermore, process parameters and the machine tool's structural resonances modulate both the amplitude and frequency content of the signal, while external disturbances can introduce forced vibrations and background noise. [6]. The simultaneous acquisition of signals from both the spindle and tailstock significantly enhances robustness by mitigating the impact of localized noise sources that might not be fully transmitted to the workpiece. Consequently, each observation is initially represented by a pair of spectrograms, which can be conceptualized as third-order tensors with modes corresponding to frequency, time, and sensor channel. To emphasize shared features and reduce sensor-specific noise, the two spectrograms are combined through elementwise (Hadamard) multiplication, defined as:

$$S_m(t,f) = S_{sp}(t,f) \circ S_{tl}(t,f) \qquad (1)$$

where $S_{sp}(t,f)$ and $S_{tl}(t,f)$ are signals acquired on the spindle and tailstock, respectively, after amplitude normalization; the operator "∘" denotes the elementwise (Hadamard) product, while $S_m(t,f)$ denotes the resulting merged map. Finally, the combined maps $S_m$ are subsequently rescaled to further enhance the identification of critical signal patterns.

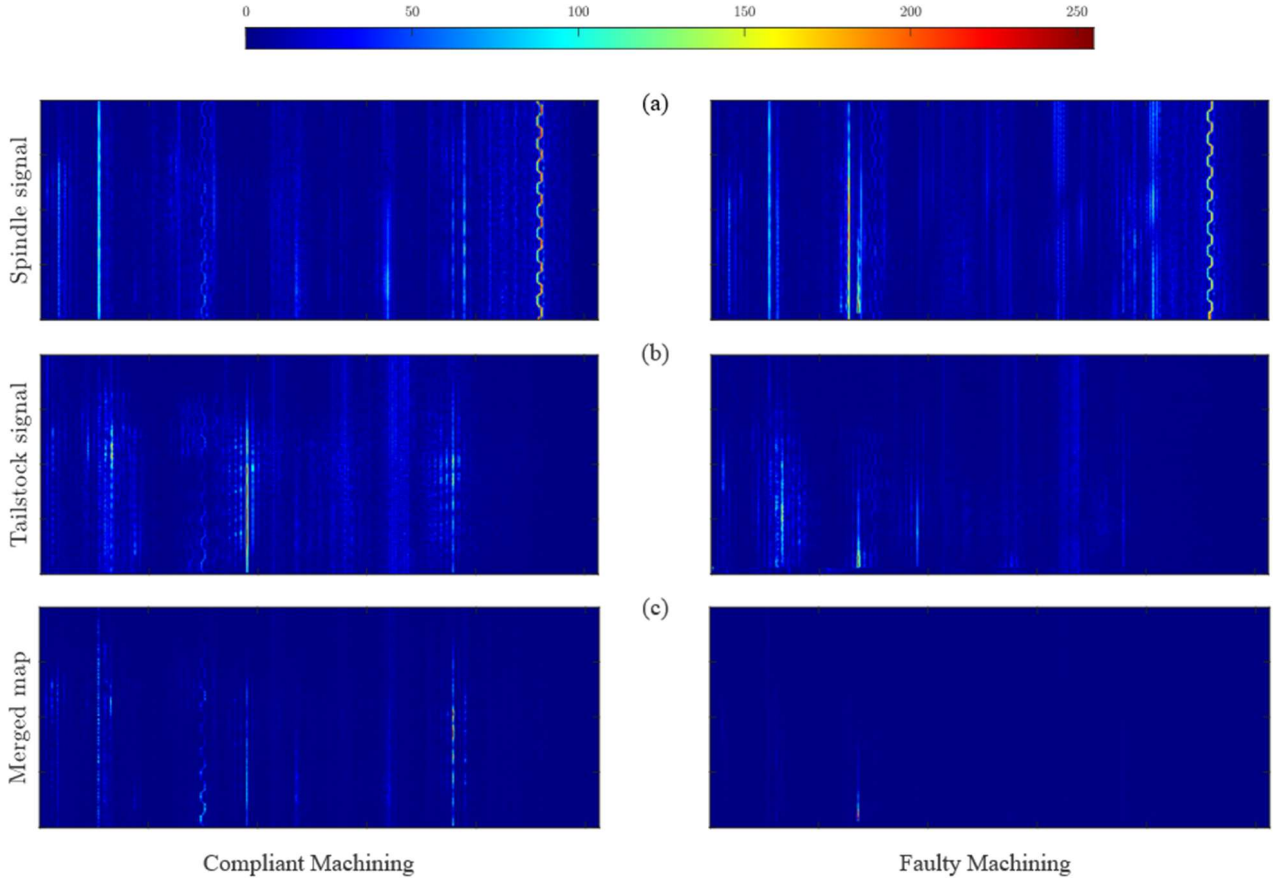

*Figure 3 - Examples of acceleration spectrograms related to compliant (on the left) and faulty machining (on the right). From the top to the bottom: signals acquired on the spindle (a) tailstock (b) and their combination (c). All the spectrograms are represented according to the grayscale [0,255].*

Figure 3 examples of individual spindle and tailstock spectrograms and their combined map, illustrating both compliant (left column) and faulty (right column) machining conditions. The merged map effectively isolates dynamic patterns strongly associated with gear surface quality and improves the visibility of common frequency components arising from the intricate interactions among the clamping system, workpiece, and centering system. For linear subspace learning approaches, such as PCA and LDA, the input data must be reshaped into a column vector (first-order tensors) for each observation, resulting in a dimensionality of $D = w \times b$.

### 2.3.2 Principal component analysis

Through PCA we aim to reduce the dimensionality of the initial representation and to uncover common patterns in the signal. Given a training set $X$, of $N$ samples expressed as $\{x_1\ x_2 \dots x_N\}$, where each observation $x_i$ lays in a *D*-dimensional space (map space), the *principal components* (PCs) $\boldsymbol{u}_p$ are the eigenvectors of the scatter matrix, corresponding to the *P* largest eigenvalues $\lambda_p$, obtained solving the *eigenvalue problem*

$$\boldsymbol{S}_T \boldsymbol{u}_p = \lambda_p \boldsymbol{u}_p, \qquad p = 1,2,\dots,P \qquad (2)$$

and the total scatter matrix of the training set is defined as:

$$\boldsymbol{S}_T = \sum_{j=1}^{N} (\boldsymbol{x}_j - \boldsymbol{\mu})(\boldsymbol{x}_j - \boldsymbol{\mu})^\mathrm{T} \qquad (3)$$

where $\boldsymbol{\mu}$ denotes the mean vector of the training set. The eigenvectors $\boldsymbol{u}_p$ are *D*-dimensional vectors and, according to Ref. [26], can be reshaped into *eigen-maps* corresponding to $(w,b)$-size image. PCA is a linear transformation which reduces the original *D*-dimensional map space into a *P*-dimensional feature space, where $P < D$. Any observation $\boldsymbol{b} = \boldsymbol{x} - \boldsymbol{\mu}$, centered with the training mean $\boldsymbol{\mu}$, can be projected onto a lower-dimensional subspace or feature space, as follows

$$\boldsymbol{y} = \boldsymbol{U}_{pca}^T \boldsymbol{b} \qquad (4)$$

where $\boldsymbol{U}_{pca}$ is the projection matrix whose columns are the principal components $\boldsymbol{u}_p$, and $\boldsymbol{y}$ is the low-dimensional representation of $\boldsymbol{b}$ in the PC space, which preserves the most relevant information shared by the input signals. Since the objective of this study is to assign the correct class label to each observation, the optimal dimensionality *P* is determined based on the classification performance evaluated on the training, cross-validation and test sets.

### 2.3.3 Linear Discriminant Analysis

When class labels are available, incorporating this information into the dimensionality reduction process can significantly improve classification performance. Supervised subspace learning methods, such as Linear Discriminant Analysis (LDA), have been shown to outperform unsupervised approaches like PCA, by optimizing the projection directions to enhance class separability, improving subsequent classification results [24]. The classical LDA is based on Fisher's discriminant criterion (FDC), which seeks a linear projection of high-dimensional data onto a lower-dimensional subspace in which the between-class scatter and within-class scatter ratio is maximized. Given a labeled training dataset $\boldsymbol{X}$ with $c$ classes, where each class contains $n_i$ set of samples $\boldsymbol{X}_i$ with $i = 1,2,\dots,c$, the *discriminant components* (DCs) $\boldsymbol{u}_p$ are obtained solving the following *generalized eigenvalue problem*:

$$\boldsymbol{S}_B \boldsymbol{u}_p = \lambda_i \boldsymbol{S}_W \boldsymbol{u}_p, \qquad p = 1,2,\dots,P \qquad (5)$$

where $\boldsymbol{S}_B$ is the between-class scatter matrix and $\boldsymbol{S}_W$ is the within-class scatter matrix, which measures the variability of training data, respectively between the different classes and within each class

$$\boldsymbol{S}_B = \sum_{i=1}^{c} n_i(\boldsymbol{\mu}_i - \boldsymbol{\mu})(\boldsymbol{\mu}_i - \boldsymbol{\mu})^\mathrm{T} \qquad \boldsymbol{S}_W = \sum_{i=1}^{c} \sum_{\substack{j=1 \\ x_j \in X_i}}^{n_i} n_i(\boldsymbol{x}_j - \boldsymbol{\mu})(\boldsymbol{x}_j - \boldsymbol{\mu})^T \qquad (6)$$

where $\boldsymbol{\mu}_i$ denotes the mean vector for each training set $\boldsymbol{X}_i$. The number of discriminants components and associated eigenvalues $\lambda_i$ is at most $P' \leq c - 1$, due to the rank of $\boldsymbol{S}_B$. To avoid the singularity of the within-class scatter matrix $\boldsymbol{S}_W$, which is common when the features dimensionality exceeds the number of samples ($D > N$), we adopt the two-stage approach described in [23]. Here, PCA is first applied to reduce the dimensionality of the input dataset from *D* to *P* by using the projection matrix $\boldsymbol{U}_{pca}$, whose columns are the $\boldsymbol{u}_p$ defined in equation (2), ensuring that $\boldsymbol{S}_W$ is full rank ($P \leq N - c$). Subsequently, LDA is applied to project the data, obtained by equation (2), from the *P*-dimensional subspace to a *P'*-dimensional discriminant subspace, by using the projection matrix $U_{lda}$ whose columns are the $u_p$ defined in equation (5). Any observation $b = x - \mu$, centered with the training mean $\mu$, can be projected onto the feature space, as follows:

$$y = \widetilde{U}^T b \tag{7}$$

where $\widetilde{U} = U_{pca} U_{lda}$ is the projection matrix, whose columns are the combined features, and $y$ is the low-dimensional representation of $b$ in the discriminant space. Let's observe that $U_{pca}$ is a $D \times P$ matrix and $U_{lda}$ is a $P \times P'$ matrix, then the resulting projection matrix $\widetilde{U}$ has dimensions $D \times P'$.

### 2.3.4 Uncorrelated Multilinear Discriminants Analysis with regularization

Uncorrelated Multilinear Discriminant Analysis (UMLDA) is a supervised multilinear subspace learning approach, an extension of classical LDA. It is specifically designed to operate directly on high-order tensors, thereby preserving the intrinsic structural relationships within the data by performing mode-wise projections rather than flattening tensors into vectors [27]. This method maximizes the scatter ratio (according to Fisher's Discriminant Criterion) while enforcing orthonormality of the extracted features, which ensures their decorrelation. Given a training set of N $M$-mode tensor samples $\{\mathcal{X}_1, \mathcal{X}_2, \ldots, \mathcal{X}_N\}$, each associated with a class label, the multilinear projections $\{u_p^{(m)^T}, m = 1,2,\ldots,M\}_{p=1}^P$ are computed iteratively using alternating projections. This involves fixing all projections except one and solving the generalized eigenvalue problem

$$S_{B_p}^{(m)} u_p^{(m)} = \lambda S_{W_p}^{(m)} u_p^{(m)}, \quad m = 1,2,\ldots,M \text{ and } p = 1,2,\ldots,P \tag{8}$$

where $S_{B_p}^{(m)}$ and $S_{W_p}^{(m)}$ are respectively the between-class scatter matrix and the within-class scatter matrix, for the $m$-th mode and $p$-th projection. Any observation $\breve{\mathcal{X}}$, centered with the mean training matrix, can be projected onto the feature space as follows:

$$y = \breve{\mathcal{X}} \times_{m=1}^M \{u_p^{(m)^T}, m = 1, \ldots, M\}_{p=1}^P \tag{9}$$

where $\times_m$ denotes the m-mode product of a tensor by a matrix, which maps the original tensor space to a lower dimensional subspace, and $y$ is the low-dimensional representation of $\mathcal{X}$. To address the singularity problem of within-class scatter matrix, which commonly occurs when the feature dimensionality $D$ exceeds the number of samples $N$, a regularization parameter $\gamma$ is incorporated as defined in Ref. [27]. A critical aspect of this study is the application of R-UMLDA under two distinct data configurations: (i) merged spectrograms, where the dimensionality $D$ of the samples is $w \times b$, considering the combined spectrograms obtained from the spindle and tailstock sensors (leading to second-order tensors); (ii) spectrograms pairs, where the input consists of two spectrograms for each observation (one from each sensors), leading to third-order tensors with dimensionality and $w \times b \times 2$. The classifier performance based on features extracted by R-UMLDA is rigorously evaluated for both these data configurations, allowing for a comprehensive assessment of the method's ability to leverage the inherent multi-dimensional structure of the data.

### 2.3.5 Support Vector Machine

In the eigen-maps subspace, data samples exhibiting similar signal patterns tend to cluster based on shared underlying features, requiring an appropriate classification criterion to accurately identify and separate the corresponding classes. For this purpose, we used a multiclass kernelized Support Vector Machine (SVM) model. SVM is a supervised machine learning algorithm widely used for industry and science tasks. The primary goal of SVM is to find the optimal hyperplane that maximizes the margin between the data points of two different classes. For *linearly* separable data, the hyperplane is defined by:

$$w \cdot x + b = 0 \tag{10}$$

where $w$ is the weight vector, $x$ the observation and $b$ the bias term. The SVM aims to maximize the margin, which is the distance between the closest data points of the two classes, also known as support vectors. The optimization problem can be expressed synthetically as:

$$\min_{w,b} \frac{1}{2}\|w\|^2, \quad y_j(wx_j + b) \geq 1 \; j = 1,2,\dots,N \tag{11}$$

where $y_j = \{\pm 1\}$ is the class label for each data point $x_j$. In general, real-world applications require non-linear functions and kernel representations offer an alternative solution by projecting the data into a high dimensional feature space to increase the computational power of the linear learning machines [31]. The idea beyond is that a representation that matches the specific learning problem should be chosen, mapping the input data $x$ into a new space [32]:

$$x \to \Phi(x) \quad f(x) = w \cdot \Phi(x) + b \tag{12}$$

where the vector $w$ is expressed by $w = \sum_{j=1}^{N} \alpha_j \Phi(x_j)$, with $\alpha_j$ weights associated to each $\Phi(x_j)$:

$$f(x) = \sum_{j=1}^{N} \alpha_j \Phi(x_j) \cdot \Phi(x) + b = \sum_{j=1}^{N} \alpha_j K(x_j, x) + b \tag{13}$$

where $K(x_j, x)$ is the kernel function. Here we employed a Gaussian kernel function, $K(x_j, x) = e^{-\|x_j - x\|^2}$.

To extend binary classifiers to multi-class classification problems, the Error-Correcting Output Codes (ECOC) technique is widely employed. This latter transforms a multi-class classification problem into multiple binary classification problems [33]. In this work, an ECOC classifier based on binary SVM model is trained based on the projections into the feature space.

## 3 Results and Discussion

This section presents and critically discusses the experimental results, focusing on the efficacy of the proposed AI-integrated methodology for in-process gear quality assessment. The primary objective is to evaluate the predictive capabilities of the different subspace learning models in accurately classifying gear quality classes based on time-frequency patterns from vibration spectrograms. Model selection and optimization involve determining the optimal subspace dimension, P, which corresponds to the number of projection features extracted by each subspace learning model. This parameter was optimized based on its impact on overall classification accuracy (correct classification rate) as a function of the number of features. To ensure robust generalization and prevent overfitting, a 5-fold cross-validation strategy was applied to the training set. Final model performance was then rigorously assessed by projecting unseen test samples into the learned feature space. To address potential class imbalance within the dataset, the F1-score was additionally computed for each class after feature selection, providing a more nuanced evaluation of predictive performance across different quality categories. Beyond quantitative performance metrics, this study also emphasizes the interpretability of the extracted features. The selected features are visualized as grayscale eigen-maps, which offer valuable insights into the distinct operating conditions and underlying physical phenomena of the machine tool. To further quantify and interpret the contribution of each eigen-map in reconstructing the spectrograms associated with the four quality classes, the average interpretation coefficient $\theta$, as defined in Ref. [26], was utilized. For the multilinear case (R-UMLDA), this coefficient was extended by employing tensor-to-vector projection instead of scalar projection. The following subsections present detailed results for each subspace learning approach, providing a comparative analysis and a discussion of their practical and scientific implications. The data distribution within the learned subspace is also graphically displayed as scatter plots to visualize class separability.

### 3.1 Features extraction by PCA

The classification accuracy of the ECOC model, utilizing features extracted via PCA, is presented in Figure 4 as a function of the number of principal components (P). The results, encompassing training, cross-validation, and test sets, indicate that the optimal number of features for PCA is five, beyond which overfitting is observed, particularly in the cross-validation curve. Consequently, all subsequent PCA-based results are derived using P

= 5 features. Table 3 and Figure 5 provide the corresponding class-wise F1-scores and confusion matrix for the test set, respectively.

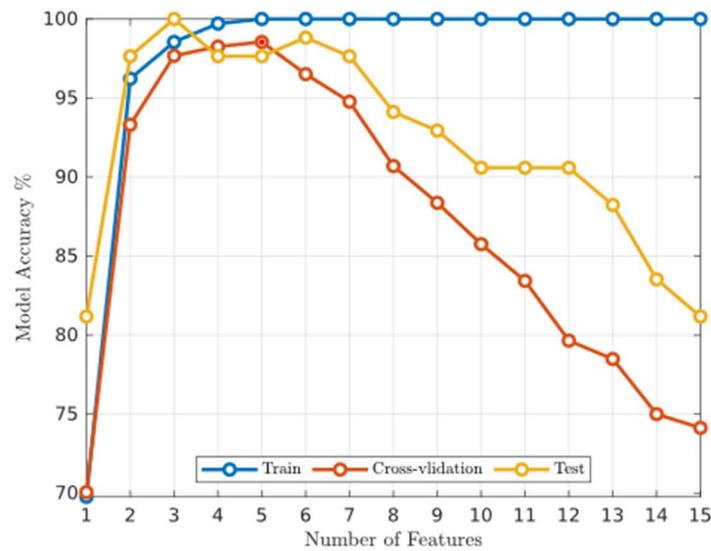

*Figure 4 – ECOC model accuracy for training, cross validation and test, as a function of the number of features extracted by PCA. The optimal PCs number is highlighted as a red dot on the cross-validation curve.*

The overall accuracy for the test set achieved with PCA is 97.65% (Table 1), demonstrating its effectiveness in dimensionality reduction and feature extraction for this task. However, a closer look at the confusion matrix (Figure 5) reveals that while classes OK, NOK 1, and NOK 2 are classified with very high accuracy (96.3% to 100% F1-score), the NOK 3 class exhibits a lower F1-score of 85.71%. This indicates that PCA, being an unsupervised method, struggles slightly with the smaller, more challenging NOK 3 class, as it maximizes total variance without explicit consideration for class separability. Specifically, one instance of NOK 3 was misclassified as NOK 1, and one instance of NOK 1 was misclassified as OK.

|  | OK | NOK 1 | NOK 2 | NOK 3 |
|---|---|---|---|---|
| Training (%) | 100.00 | 100.00 | 100.00 | 100.00 |
| Cross-validation (%) | 98.64 | 98.19 | 99.54 | 100.00 |
| Testing (%) | 96.30 | 96.18 | 100.00 | 85.71 |

*Table 1 - F1-score by PCA analysis.*

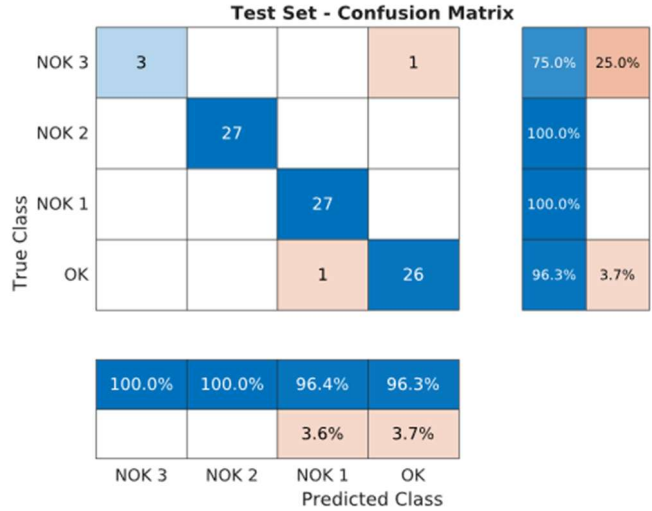

*Figure 5 – Test set confusion matrix.*

The first five eigen-maps, derived from solving Equation (2) and reshaped to the original spectrogram dimensions ($w \times b$.), are visualized in grayscale in Figure 6. These eigen-maps define a subspace of uncorrelated coordinates, enabling a synthetic representation of the original dataset. Figure 7 illustrates the average interpretation coefficient, $\theta$, for each class on the test samples, quantifying the contribution of each of the first five eigen-maps to spectrogram reconstruction. The data distribution in this feature space is depicted in Figure 8, showing projections onto the first three PCs (left) and the most discriminative PCs (right).

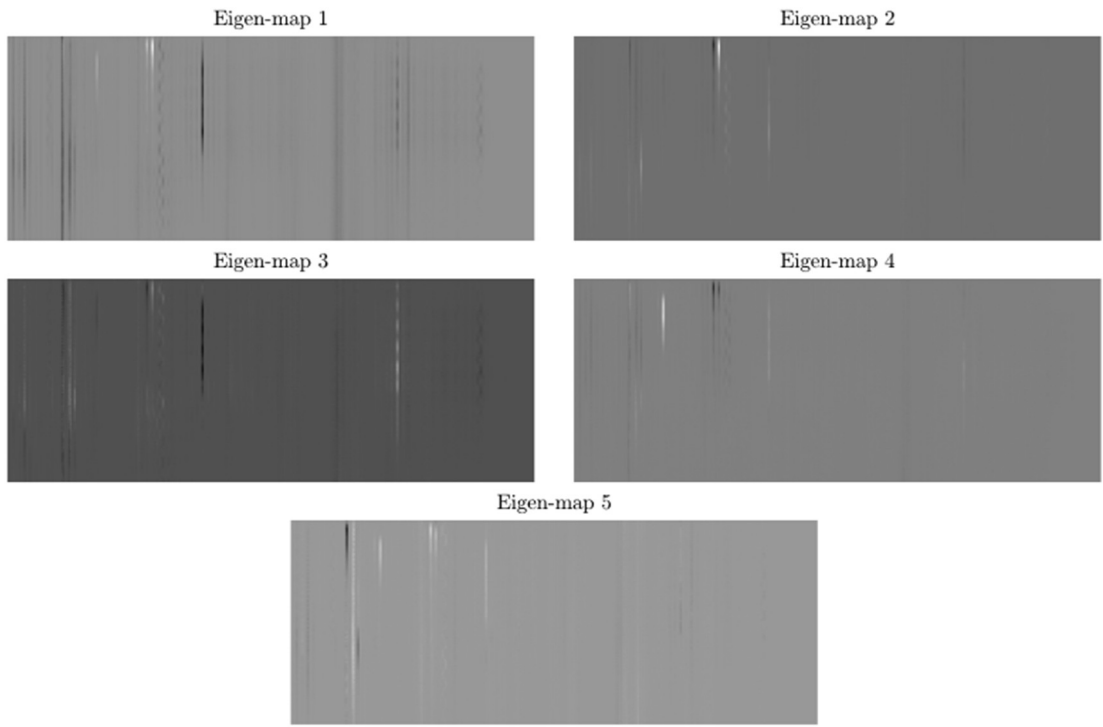

*Figure 6 - Grayscale representation of the first five eigen-maps.*

The analysis of the eigen-maps (Figure 6) and their interpretation coefficients (Figure 7) provides critical insights into the underlying process dynamics and the nature of gear defects. Specifically, Eigen-map PC 1 exhibits a substantial contribution across all classes, with a particularly strong influence on the 'OK' (compliant) class. As observed in the data distribution (Figure 8), PC 1 effectively separates the 'OK' class samples from all non-compliant ('NOK') ones. Practically, this suggests that PC 1 captures the general "healthy" state of the honing process, reflecting the predominant vibration patterns associated with properly manufactured gears. Eigen-map PC 2 shows a relevant contribution specifically for classes NOK 1 and NOK 2, enabling a clear distinction between these two non-compliant categories. This implies that PC 2 is sensitive to the specific vibration characteristics that differentiate between NVH performance issues (NOK 1, linked to surface waviness) and micro-geometry deviations (NOK 2). Eigen-maps PC 3 and PC 4 show limited contribution to spectrogram reconstruction and class discrimination, hence they do not form distinct clusters. From a practical standpoint, this confirms that these PCs capture less relevant variance for defect classification, supporting the choice of a lower number of features. Finally, eigen-map PC 5 makes the largest contribution to the NOK 3 class, effectively separating it from all other classes. This is particularly significant as NOK 3 corresponds to NVH issues due to periodic pitch patterns, suggesting that PC 5 captures unique vibrational signatures associated with this specific and challenging defect type.

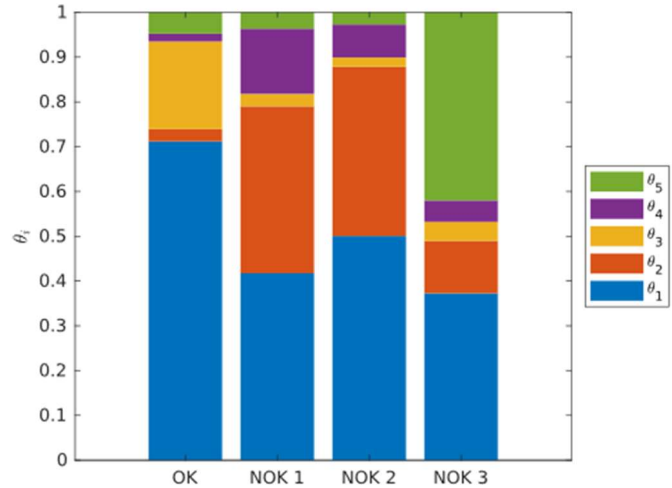

*Figure 7 – Class-wise mean interpretation coefficient $\theta$.*

The effective clustering of data in the PCs subspace (Figure 8) confirms that PCA successfully extracts features that capture the dominant variance in the dataset, which, in turn, correlates with different gear quality states. While effective, the unsupervised nature of PCA means it prioritizes total variance, which might not always align perfectly with class separability, especially for highly imbalanced or subtle defect classes. This observation motivates the exploration of supervised subspace learning methods in the subsequent sections.

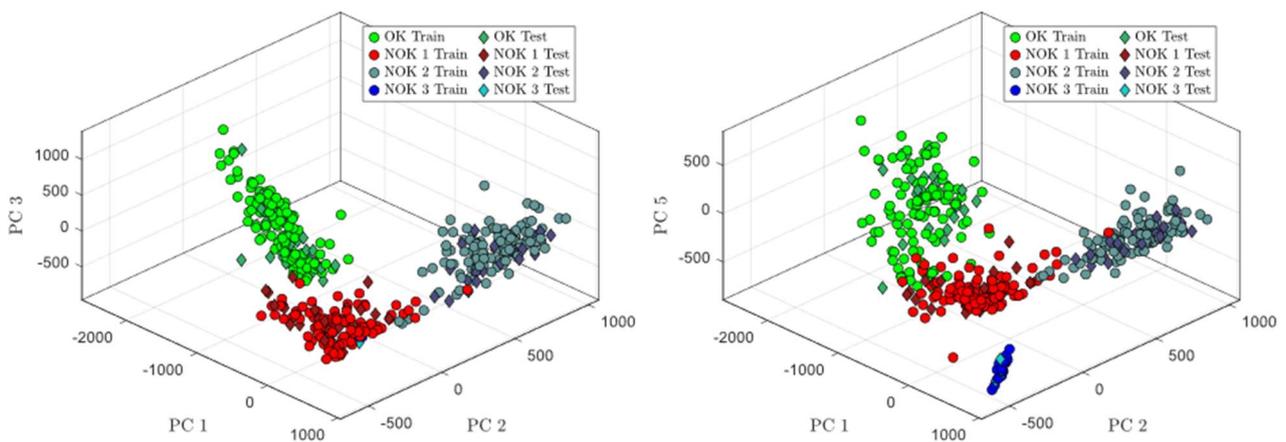

*Figure 8 - Observations in the subspace defined by the first three PCs on the left and by the $1^{st}, 2^{nd}$ and $5^{th}$ PCs on the right.*

## 3.2 Features extraction by LDA combined with PCA

Building upon the insights gained from PCA, this section evaluates the performance of a supervised two-stage subspace learning approach, combining PCA for initial dimensionality reduction with Linear Discriminant Analysis (LDA) for enhanced class separability. Figure 9 illustrates the classification accuracy of the ECOC model as a function of the number of features extracted using this hybrid approach, presenting results for the training, cross-validation, and test sets.

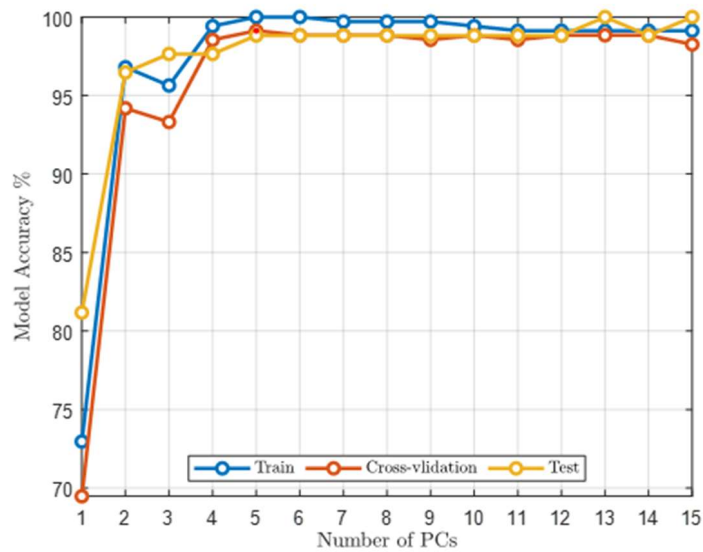

*Figure 9 - ECOC model accuracy for training, cross validation and test, as a function of the number of features extracted by PCA, used as input for LDA. The optimal PCs number is highlighted as a red dot on the cross-validation curve.*

The optimal classification performance is achieved with five features, as indicated by the cross-validation trend. Subsequent results are based on the first $P = 5$ features extracted by PCA, projected on $P' = 3$ discriminant components. Table 4 and Figure 10 present the class-wise F1-scores and the confusion matrix for the test set, respectively. The overall classification accuracy for the test set achieved with the PCA-LDA approach is 98.83% (Table 4).

|  | OK | NOK 1 | NOK 2 | NOK 3 |
|---|---|---|---|---|
| Training (%) | 100.00 | 100.00 | 100.00 | 100.00 |
| Cross-validation (%) | 99.54 | 98.62 | 99.09 | 100 |
| Testing (%) | 98.11 | 98.18 | 100.00 | 100.00 |

*Table 2 - F1-score by PCA-LDA analysis.*

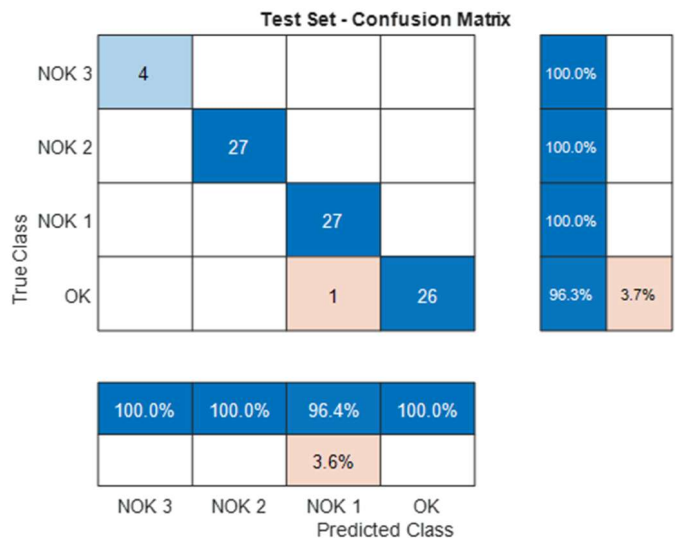

*Figure 10 - Test set confusion matrix.*

This represents a notable improvement over the standalone PCA method (97.65%), particularly evident in the F1-scores for the challenging NOK 3 class. The confusion matrix (Figure 10) confirms that only one instance of NOK 1 was misclassified as OK, while all other classes achieved 100% classification accuracy. This superior performance underscores the advantage of incorporating class-label information during feature extraction, allowing LDA to find projections that explicitly maximize between-class variance while minimizing within-class variance.

The first three Discriminant Components (DCs), or eigen-maps, derived from solving Equation (5) and reshaped to the original spectrogram dimensions ($w \times b$), are visualized in grayscale in Figure 11. Figure 12 presents the average interpretation coefficient, $\theta$, for each class on the test samples, detailing the contribution of these three eigen-maps to spectrogram reconstruction. The data distribution within the resulting feature space is graphically displayed in Figure 13, showing projections onto the first three DCs subspace.

The first eigen-map DC 1, similar to PC 1, shows a significant contribution across all classes, with a particularly strong influence on the OK class. Visually, from the data distribution in Figure 13, DC 1 effectively separates the OK samples from all NOK ones. This suggests that DC 1 captures the fundamental 'healthy' vibration patterns of the process, and LDA refines this separation by emphasizing the differences between OK and NOK states. The second eigen-map DC 2 exhibits a relevant contribution primarily for classes NOK 1 and NOK 2, enabling their clear distinction. The enhanced separation of these two classes compared to PCA highlights LDA's effectiveness in isolating the distinct vibrational signatures associated with NVH performance issues (NOK 1, due to surface waviness) and micro-geometry deviations (NOK 2). This provides a more robust diagnostic capability for these critical defect types. Finally, the third eigen-map DC 3 shows the largest contribution to the NOK 3 class, successfully separating it from all other classes. The marked improvement in the F1-score for NOK 3, combined with the strong contribution of DC 3 to this class, indicates that LDA is highly effective at identifying the subtle and unique vibration characteristics associated with periodic pitch patterns.

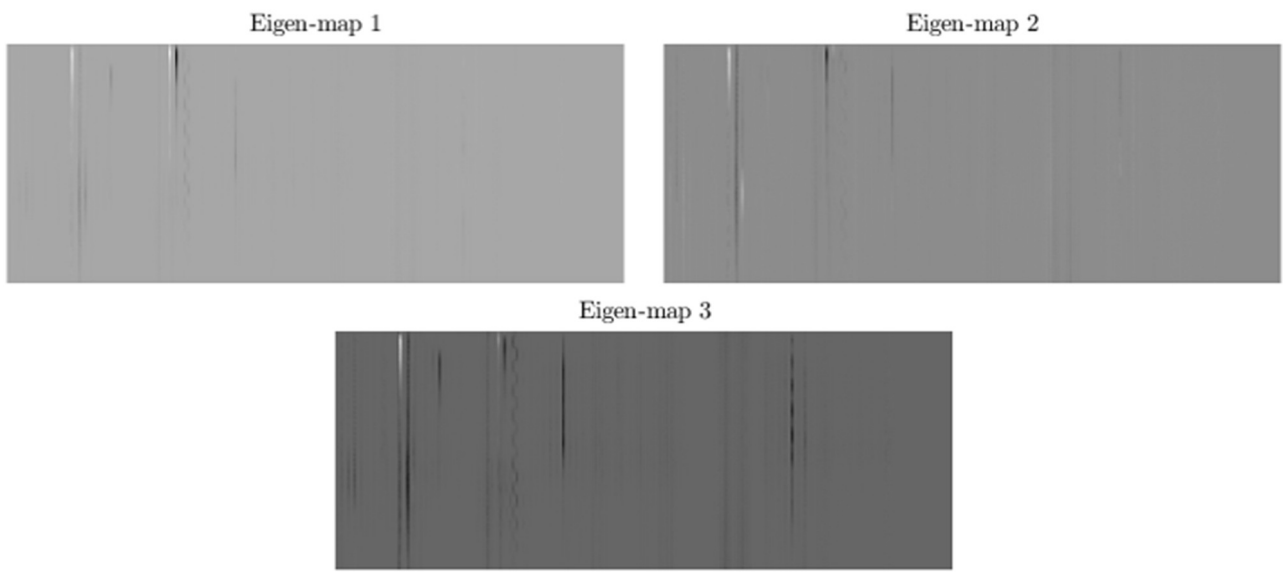

*Figure 11 - Grayscale representation of the three combined eigen-maps obtained by PCA and LDA.*

The visually clear clustering of data in the discriminant subspace (Figure 13) confirms that the PCA-LDA combination successfully extracts features that not only reduce dimensionality but also actively enhance class separability. The supervised nature of LDA proves beneficial, especially for distinguishing between different types of non-conformities, offering a more refined and diagnostically valuable feature set compared to unsupervised PCA.

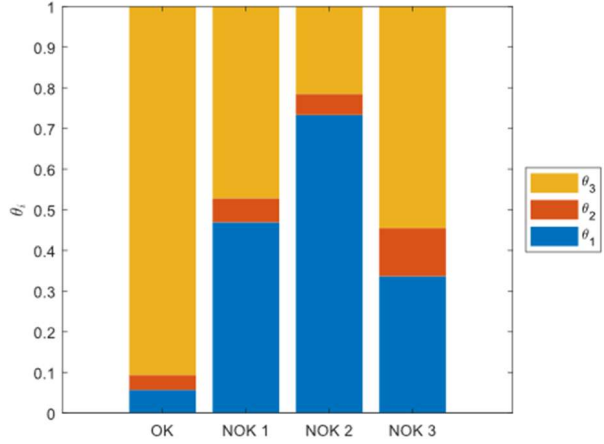

*Figure 12 – Class-wise mean interpretation coefficient.*

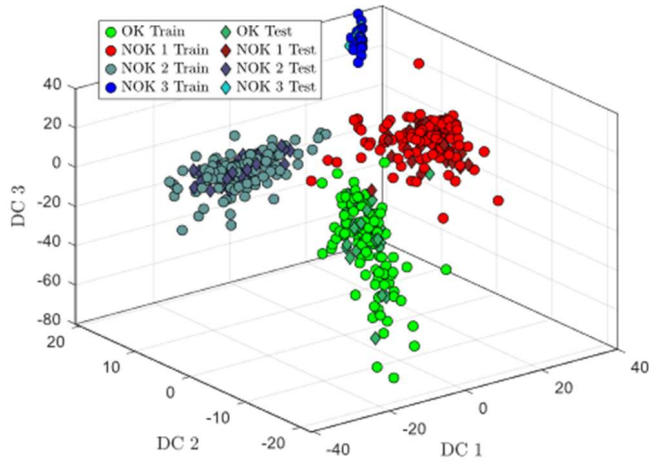

*Figure 13 - Observations in the subspace defined by the DCs.*

## 3.3 Features extraction by R-UMLDA

This section explores the performance of Uncorrelated Multilinear Discriminant Analysis with Regularization (R-UMLDA), a method specifically designed to handle high-order tensor data, thereby preserving the inherent multi-dimensional structure of the spectrograms. R-UMLDA was evaluated under two distinct input configurations: (1) using the merged 2D spectrograms (second-order tensors), and (2) using the raw spectrogram pairs from both sensors as third-order tensors (frequency VS time VS sensor channel). Comparative analysis between the two configurations revealed no significant performance differences, consequently, we reported only the results for the merged spectrogram configuration. The optimal regularization parameter is evaluated in the range $[10^{-5}, 10^{-1}]$, and the best overall results in terms of classification accuracy are obtained for $\gamma = 10^{-2}$.

Figure 14 displays the classification accuracy of the ECOC model using features extracted by R-UMLDA on the merged spectrograms. The analysis of training, cross-validation, and test sets reveals that an optimal number of three features ($P = 3$) yields the best performance without overfitting. The corresponding F1-scores for each class are reported in Table 3, and the associated the confusion matrix is omitted since 100% classification accuracy is achieved (Table 4). This represents significant advancement over both PCA and PCA-LDA, as evidenced by the perfect F1-scores across all four quality classes, including the challenging NOK 3 class.

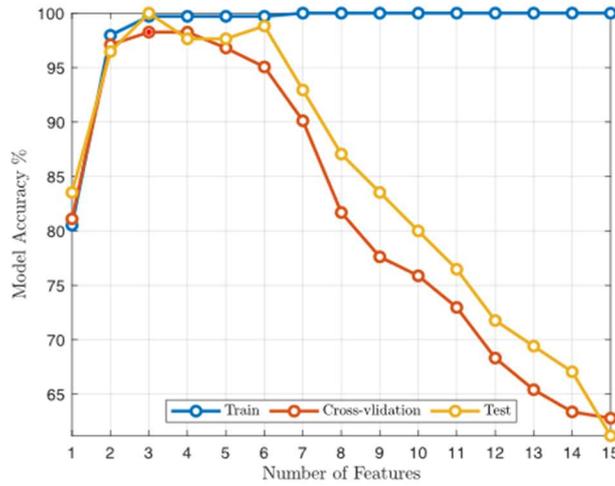
*Figure 14 - ECOC model accuracy as a function of the number of features extracted by R-UMLDA, using merged maps as input.*

|  | OK | NOK 1 | NOK 2 | NOK 3 |
|---|---|---|---|---|
| Training (%) | 100.00 | 99.54 | 99.54 | 100.00 |
| Cross-validation (%) | 99.54 | 97.74 | 98.15 | 100.00 |
| Testing (%) | 100.00 | 100.00 | 100.00 | 100.00 |

*Table 3 - F1-score by R-UMLDA analysis on merged spectrogram.*

The first three features, or eigen-maps, derived from solving Equations (8) along the frequency and time modes are displayed in the grayscale in Figure 15, while the class-wise mean interpretation coefficient $\theta$ associated with the three eigen-maps and the data distribution onto the learned subspace are presented in Figure 16 e 17, respectively.

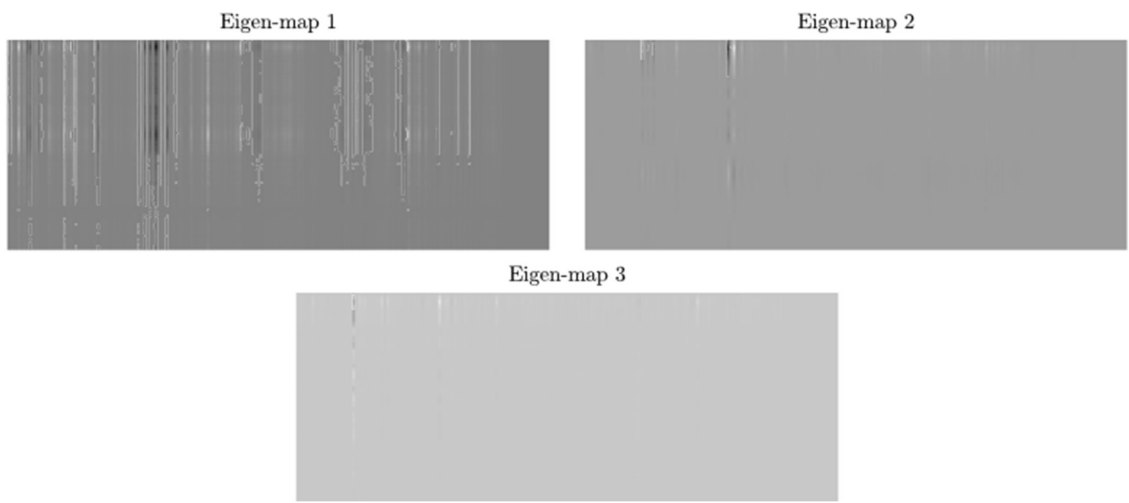

*Figure 15 – Grayscale representation of the first three eigen-maps obtained by R-UMLDA.*

The first eigen-map (Feature 1) contributed almost 100% for class OK, and cluster this latter samples apart from all the NOK ones. Eigen-map DC 1 again shows a dominant contribution across all classes, with a particularly strong influence on the 'OK' class, effectively separating it from all 'NOK' classes. The second eigen-map DC 2 (Feature 2) exhibits strong relevance for classes NOK 1 and NOK 2, enabling a precise distinction between them. Therefore, R-UMLDA is highly sensitive to the subtle differences in vibration patterns associated with NVH issues (NOK 1) versus micro-geometry deviations (NOK 2). Eigen-map DC 3 exhibits the strongest contribution to the NOK 3 class, achieving perfect separation. This is perhaps the most impactful result, as NOK 3 (NVH issues from periodic pitch patterns) often presents with very subtle signatures.

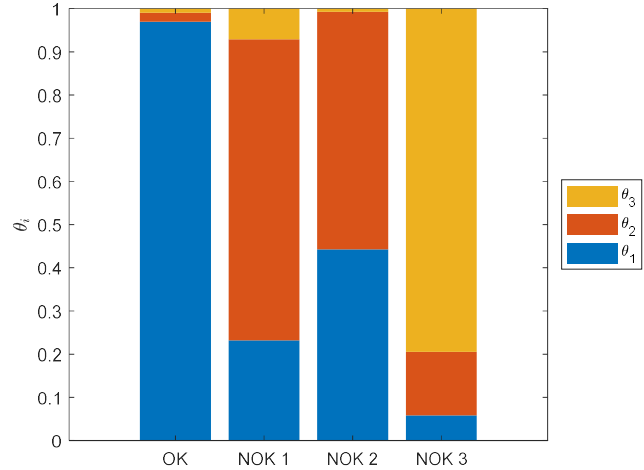

*Figure 16 - Class-wise mean interpretation coefficient.*

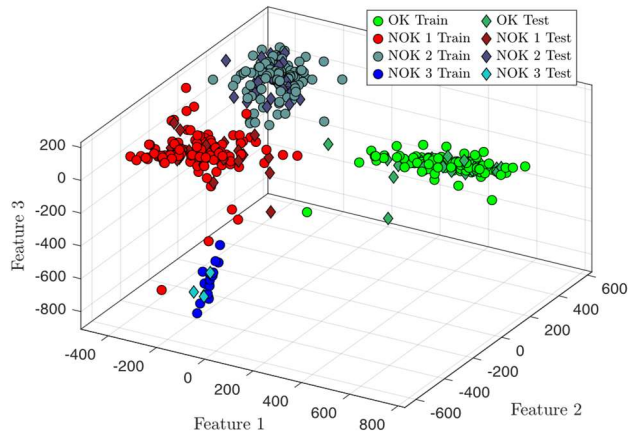

*Figure 17 - Observations in the subspace defined by R-UMLDA features.*

The consistently clear clustering of data in the discriminant subspace (Figure 17) reinforces the effectiveness of R-UMLDA in extracting features that are optimally discriminative. The perfect classification accuracy, coupled with the interpretable nature of the eigen-maps, indicates that R-UMLDA, even when applied to merged 2D spectrograms, leverages the data's structure more effectively for supervised classification.

## 3.4 Comparative Analysis of Models

A comprehensive comparison of the subspace learning approaches, along with the corresponding ECOC classification accuracies, is presented in Table 7. Let's observe that among all, PCA requires more than three features to achieve a stable learning curve, due to its unsupervised nature, based on maximizing total variance without leveraging class label information. In contrast, the supervised methods, LDA combined with PCA and R-UMLDA approach achieve high accuracy with only three features, highlighting the benefit of incorporating class-discriminative information into the projection process. The best classification performance (100% accuracy on the test set) is obtained using features extracted via R-UMLDA applied to the merged spectrograms. Although R-UMLDA does not require flattening of multidimensional data, it involves a more computationally intensive training phase, due to the alternating optimization process required for tensor projections and regularization parameter selection. Classical linear subspace learning methods, offer computational efficiency and flexibility, however they present intrinsic limitations, primarily due to the requirement of data flattening into vectors before projection, which can break the natural structure of the original data and introduces additional preprocessing steps in the monitoring. From a practical perspective, the use of a reduced number of projection features, combined with lightweight signal preprocessing, translates into greater computational efficiency during the process monitoring stage, an essential requirement in real-time industrial applications.

|  | PCA | PCA - LDA | R-UMLDA |
|---|---|---|---|
| Input data | Flattened maps | | Maps |
| Features number | 5 | 3 | 3 |
| Classification accuracy on test set (%) | 97.65 | 98.82 | 100.00 |

*Table 4 – Comparison of feature extraction methods and corresponding ECOC classification results.*

# 4 Conclusions

Characterizing the non-stationary phenomena associated with surface waviness on gear flanks—which result in unacceptable vibrations during gear operation—requires capturing the full temporal and spectral content of vibration signals. This work proposes a novel methodology that leverages time–frequency spectrograms and subspace learning to extract compact, discriminative, and interpretable features suitable for in-process monitoring of gear finishing operations.

Three subspace learning methods (PCA, PCA–LDA, and R-UMLDA) were explored to reduce the high dimensionality of spectrograms while preserving the most relevant information. The features obtained were used to train SVM classifiers for anomaly detection, with a focus on identifying noisy gears. Results reveal that a small number of features carry the most discriminative power, while adding more leads to overfitting and introduces noise. This confirms the importance of low-rank representations for robust generalization in real-world settings.

A key advantage of the proposed framework lies in its low computational cost, which makes it highly suitable for real-time applications where processing time is constrained. The extracted features—referred to as eigen-maps—offer meaningful and compact representations of the spectrograms, enabling accurate gear quality classification. An interpretability coefficient was also introduced to quantify the relevance of each feature to the classification task, enhancing transparency and trust in the system's outputs.

Beyond classification, the reduced dimensionality and discriminative strength of the features make the framework ideal for trend analysis and predictive maintenance integration. Future work will investigate the fusion of additional sensor data—such as honing wheel signals, electrical consumption, temperature, and acoustic emissions—to increase detection sensitivity and system robustness across a wider range of defects.

These findings are aligned with previous research on grinding process dynamics and advance the state of the art in data-driven condition monitoring. From a practical standpoint, the proposed method is both computationally efficient and flexible, offering a scalable solution for real-time industrial monitoring and adaptive quality control.